\documentclass[final,5p,times,twocolumn]{elsarticle}

\usepackage{amsmath}
\usepackage{wasysym}
\usepackage{graphicx}%
\usepackage{amssymb}

\def\Rb{$^{87}\mathrm{Rb}\;$}
\def\Na{$^{23}\mathrm{Na}\;$}
\def\Cs{$^{133}\mathrm{Cs}\;$}
\def\Cr{$^{52}\mathrm{Cr}\;$}
\def\mK{\;\mathrm{mK}}
\def\uK{\;\mu \mathrm{K}}
\def\nK{\;\mathrm{nK}}

\def\um{\;\mu\mathrm{m}}
\def\Hz{\;\mathrm{Hz}}
\def\kHz{\;\mathrm{kHz}}
\def\MHz{\;\mathrm{MHz}}

\def\ms{\;\mathrm{ms}}

\def\s{\;\mathrm{s}}

\def\Bz{\;\mathrm{G}/\mathrm{cm}}

\def\uWcm{\;\mu \mathrm{W}/\mathrm{cm}^2}
\def\dens{\;\mathrm{cm}^{-3}}
\def\E#1{\times 10^{#1}}
\def\rad{2\pi\times}

\journal{Optics Communications}

\begin{document}
\begin{frontmatter}

\title{All-optical Bose-Einstein condensation in a 1.06 $\mu$m dipole trap}
\author{K. J. Arnold}
\author{M. D. Barrett}
\address{Centre for Quantum Technologies/Department of Physics, National University of Singapore, 3 Science Drive 2, Singapore 117543}
\ead{cqtbmd@nus.edu.sg}

\begin{abstract}

We report the all-optical production of a \Rb Bose-Einstein condensate (BEC) in a simple $1.06\um$ dipole trap experiment. We load a single beam dipole trap directly from a magneto-optic trap (MOT). After evaporation in the single beam, a second crossed beam is used for compression. The intensity in both beams is then reduced for evaporation to BEC. We obtain a BEC with $3.5\E4$ atoms after 3 s of total evaporation time.  We also give a detailed account of the thermal distribution in cross beam traps.  This account highlights the possible difficulties in using shorter wavelength lasers to condense all optically. 

\end{abstract}

\end{frontmatter}

\section{Introduction}
Since the first observation of Bose-Einstein condensation (BEC) in 1995 \cite{anderson1995,davis1995}, many experimental groups worldwide have produced and investigated BEC. The primary method used has been radio frequency (RF) induced evaporation in magnetic traps. However, in 2001 it was demonstrated that BEC could be obtained by all optical means \cite{barrett2001}. Since then, several atomic species have been successfully condensed via this method.  With a near $10\um$ CO$_2$ laser, \Rb\cite{barrett2001,cennini2003}, \Cs\cite{weber2003}, and \Na\cite{dumke2006} have all been condensed.  With a near $1\um$ wavelength fiber laser, \Rb \cite{kinoshita2005,couvert2008,clement2009}, \Cs\cite{hung2008}, and \Cr\cite{beaufils2008} have been condensed.

The advantages of the all-optical approach over magnetic trapping include a relatively simple experimental setup, comparatively high repetition rates, and the ability to trap arbitrary spin states. The primary difficulty is associated with relaxing the trapping potential to induce evaporation. Relaxation of the trap leads to a continual decrease in the collision rate as the evaporation progresses.  The decreasing collision rate can ultimately stagnate the evaporation process before degeneracy is achieved. In Refs. \cite{barrett2001,cennini2003} researchers were able to achieve sufficiently high initial phase space and spatial densities such that BEC could be achieved before the evaporation process stagnated. Other experiments have used methods of varying complexity to circumvent slowing of the evaporation.  In Ref. \cite{kinoshita2005} a mobile lens system was used to compress the atoms to offset the decreasing density. In other experiments evaporation was forced without relaxing the optical trap using either a strong magnetic field gradient \cite{hung2008}, or a large displaced auxiliary beam \cite{clement2009}. 

The experiment reported in this paper uses a simpler approach which achieves a higher gain in the phase space density of $10^5$ than is typically achieved in other all-optical experiments. Before giving a detailed account of our experiment, we  briefly discuss the nature of a thermal distribution in bi-modal trapping potentials.  This discussion highlights the importance of the trapping wavelength in the cross beam geometry and motivates our approach to evaporation in an optical trap formed by a near $1\um$ wavelength fiber laser.  Although our discussion focuses on crossed beam dipole traps, the considerations presented also apply to composite traps with disparate volumes.  In particular, these considerations will be important when designing bichromatic optical dipole force traps for reaching Fermi degeneracy as proposed in \cite{onofrio2002}.  Such traps typically require the use of shorter wavelength trapping beams and here we highlight the specific problems associated with implementing such traps.

\begin{figure*}[t]
\centering
\includegraphics{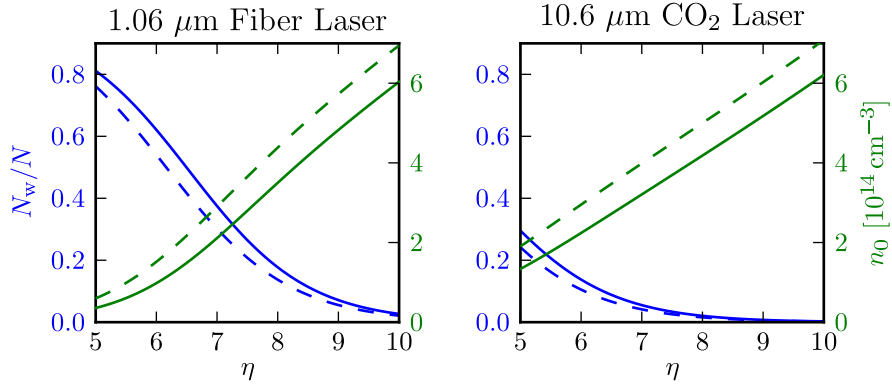}
\caption{Comparision of the fraction of atoms in the wings outside the dimple (blue lines) and the peak density (green lines) as a function of $\eta$ for $1.06\um$ and $10.6\um$ traps. The green and blue dashed lines are the analytic approximations from equations (\ref{eq:R}) and (\ref{eq:n0}) respectively. The solid lines result from numerical integration of the exact equations (\ref{eq:Ncintg}) and (\ref{eq:Nwintg}) with $\beta=1.9$.  All calculations assume an atom number $N=2\E{6}$ and beam waists $w_0=40\um$.}
\label{wingplot}
\end{figure*}

\section{Thermalization in Crossed Beam Traps}
\label{discussion}
Laser cooling into a cross beam trap results in densities greatly exceeding those typically obtained in other experiments \cite{adams1995}. To some extent this can be understood as follows.  During loading, a large number of atoms can be trapped due to the large volume associated with the two single focus beams making up the trap. After loading, the atoms begin to relax via collisions to a Boltzmann distribution appropriate to the trap geometry.  This distribution tends to have the atoms localized to the much smaller volume associated with the intersection of the beams and this results in a substantial increase in both the spatial and phase space densities. There are two crucial requirements for the success of this approach; a) the majority of atoms must indeed be localized to the intersection region and b) any atoms remaining in the wings should be in thermal equilibrium with the atoms in the center.  As the divergence of each beam decreases, the volume of the wings increases relative to the intersection region.  This increases the number of atoms in the wings in accordance with a Boltzmann distribution.  More importantly, the axial frequency associated with a single beam decreases, increasing the time atoms spend outside the intersection region and slowing the thermalization process.  Thus the wavelength of the trapping beams can have a significant impact on the thermalization process.

To make the preceding arguments more concrete we follow the methods given in Ref. \cite{mythesis}.  For finite depth traps, the trapped atoms are well characterized by a quasi thermal equilibrium in which the phase space is restricted to regions in which the total energy is less than the trap depth, $\epsilon_t$.  This leads to a density distribution given by
\begin{equation}
n(\mathbf{r})=n_0 \exp\left(-\frac{U(\mathbf{r})}{k_b T}\right)P\left(3/2,\frac{\epsilon_t-U(\mathbf{r})}{k_bT}\right)
\end{equation}
where $P(3/2,x)$ is the incomplete gamma function and $n_0 P(3/2,\epsilon_t/k_b T) \approx n_0$ is the peak density. To find $n_0$ we can integrate over all space and equate the result to the total number of atoms.  To this end it is convenient to introduce a function $V(u)$ which is the volume of space enclosed by the equipotential surface $U_0 u$. $U_0$ is a scale factor for the energy which we conveniently choose to be the depth of single beam trap. We then have
\begin{equation}
N=n_0 \int_0^\beta \exp\left(-\eta u\right)P\left(3/2,\eta(\beta-u)\right)\frac{dV}{du}d\,u
\end{equation}
where $\eta=U_0/k_b T$ and $\beta=\epsilon_t/U_0$.  As discussed in \cite{mythesis}, it is necessary to take $\beta<2$ due to a divergence in the density of states at the top of the trap.  For relevant values of $\eta$ the divergence only becomes numerically significant for $\beta \gtrsim 1.95$ and in practice the trap depth is typically less than this due to external effects such as gravity.  For $1.85 < \beta < 1.95$ the integrals only vary by $10\%$ and smaller values of $\beta$ are only relevant when the potential is significantly modified by an external influence.  Here we consider an analytic approximation and do not concern ourselves further with the choice of $\beta$.

\begin{figure}
\includegraphics{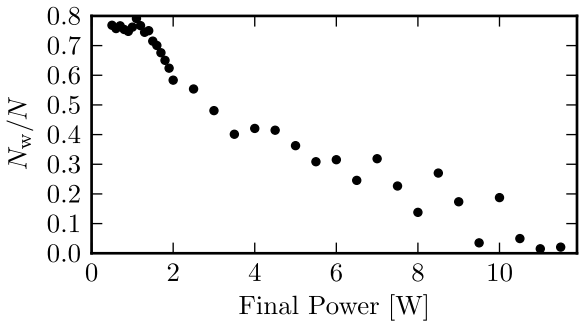}
\caption{Fraction of atoms in the wings versus final optical power after 2 s of evaporation. Slowing thermalization time in the wings leads to an increase in $\eta$ and further migration of atoms to the wings as the evaporation progresses.}
\label{wingevap}
\end{figure}

To obtain an analytic estimate of the integral, we consider the two regions $u<1$ and $u>1$ corresponding respectively to the number of atoms in the center of the trap, $N_{\mathrm{c}}$, and the number of atoms in the wings, $N_{\mathrm{w}}$. For $u<1$, the function $V(u)$ is, to a good approximation, independent of the divergence of the beams and therefore scales as $w_0^3$ where $w_0$ is the waist of the beams at the focus. Moreover, since the integral is weighted towards lower values of $u$, we make a harmonic approximation to the potential and neglect the effects of truncation.  Thus we have
\begin{eqnarray}
\label{eq:Ncintg}
N_{\mathrm{c}}&=&n_0 \int_0^1 \exp\left(-\eta u\right)P\left(3/2,\eta(\beta-u)\right)\frac{dV}{du}d\,u\\
 &\approx& \frac{\pi n_0 w_0^3}{2} \int_0^\infty \exp\left(-\eta u\right)u^{1/2}d\,u =\frac{n_0 w_0^3}{4}\left(\frac{\pi}{\eta}\right)^{3/2}.
\end{eqnarray}
For $u>1$ we have
\begin{align}
\label{eq:Nwintg}
N_{\mathrm{w}}=&n_0 \int_1^\beta \exp\left(-\eta u\right)P\left(3/2,\eta(\beta-u)\right)\frac{dV}{du}d\,u\\
\begin{split}
=&n_0 e^{-\eta} \int_0^{\beta-1}\exp\left(-\eta u^\prime \right)\\
&\quad\quad P\left(3/2,\eta(\beta-1-u^\prime)\right)\frac{dV}{du^\prime}d\,u^\prime
\end{split}
\end{align}
where we have made the substitution $u^\prime=u-1$.  For $u>1$ the function $V(u)$ is, to a good approximation, independent of the beam intersection and can be approximated by the sum of two independent single focus beams.  As before, they can be approximated by two infinitely deep harmonic oscillator pontentials giving
\begin{equation}
N_{\mathrm{w}}=n_0 w_0^2 z_R \left(\frac{\pi}{\eta}\right)^{3/2}e^{-\eta}.
\end{equation}
The fraction of atoms in the wings of the cross beam trap, $\frac{N_{\mathrm{w}}}{N}$, and the peak density at the center, $n_0$, are then easily found and we have
\begin{equation}
\label{eq:R}
\frac{N_{\mathrm{w}}}{N}\approx \frac{\frac{4\pi w_0}{\lambda} e^{-\eta}}{1+\frac{4\pi w_0}{\lambda} e^{-\eta}}
\end{equation}
and
\begin{equation}
\label{eq:n0}
n_0\approx \frac{4N}{w_0^3}\left(\frac{\eta}{\pi}\right)^{3/2}\frac{1}{1+\frac{4\pi w_0}{\lambda} e^{-\eta}}
\end{equation}
where $\lambda$ is the wavelength of the trapping beams. These expressions clearly highlight the influence of the wavelength $\lambda$.  As $\pi w_0/\lambda$ increases, a larger fraction of the atoms appear in the wings of the potential and the peak density decreases accordingly.  Figure~\ref{wingplot} illustrates the differences between a CO$_2$ laser with $\lambda=10.6\um$ and a fiber laser with $\lambda=1.06\um$.  Under typical experimental conditions, evaporation stagnates with $\eta\sim 8$ due to the very large densities and corresponding collision rates.  For both wavelengths there is only a small fraction of atoms in the wings at this value of $\eta$.  However, during forced evaporation, the collision rate drops which can result in a slight decrease in $\eta$.  Although this drop in $\eta$ is inconsequential for a CO$_2$ laser trap, it can have a substantial effect for a $1.06\um$ laser trap with large number of atoms migrating to the wings.   This was experimentally observed in our earlier attempts at evaporation as illustrated in figure~\ref{wingevap}.  Although this migration of atoms to the wings does result in a further decrease in the density, the main impact is on the thermalization and evaporation rates of the overall distribution.

To estimate the thermalization time we note that the density in the wings of the distribution is reduced by the factor $e^{-\eta}$ which substantially reduces the collision rate in that region.  Under typical experimental conditions the collision rate in the wings of the distribution is on the order of the axial frequency of the individual beams or less.  In contrast, the collision rate at the center of the trap is much higher and typically exceeds the trap frequency.  Thus, a reasonable approximation is to assume the atoms move independently in the wings until they reach the center of the trap where they have a high probability of undergoing a collision. The thermalization rate will therefore be roughly determined by the axial frequency of the individual beams.  Since this frequency scales as $\lambda/w_0^3$ the resulting thermalization time will increase by an order of magnitude in going from a $10.6\um$ CO$_2$ laser to a $1.06\um$ fiber laser.  One could consider reducing the waist to compensate, but this would increase the peak density by the same order of magnitude.  However, as illustrated in figure~\ref{wingplot}, peak densities at $w_0=40\um$ already exceed $10^{14}\dens$ where three-body losses become significant~\cite{burt1997}.  Increasing the density further would simply result in a substantial loss of atoms by this mechanism.

To avoid the problems discussed above, we employ a simple modification of the evaporation cycle.  Atoms are first loaded into a tightly focused single beam trap, and the cross beam geometry is only used later in the evaporation cycle to compress the cloud and compensate the decreasing collision rate. Using a single beam for loading provides a large number of atoms and the tight focus results in a sufficiently high density for a rapid initial evaporation. The single beam also forms part of the cross beam geometry where the tight focus ensures the atoms are well localized to the intersection region after the cloud is compressed.  During the evaporation, the density peaks after the compression but is still $<1\E{14}\dens$, so significant three-body losses are avoided.

\section{Experiment Description and Results}
\label{results}

  Our experiment consists of two $1.06\um$ wavelength trapping beams intersected at a magneto-optical trap (MOT). A standard 3D MOT is loaded from a rubidium dispenser in the experiment chamber that is run continuously. The MOT typically has $\approx 3\E8$ atoms.  The primary dipole trap is horizontal with 15~W maximum power and is focused to a $25\um$ waist.  The auxiliary crossed beam is vertical with a maximum power of 1~W and is focused to an elliptical spot with waists $20\um$ and $80\um$. The $80\um$ waist is aligned to the axis of the primary beam.

\begin{figure}
\includegraphics{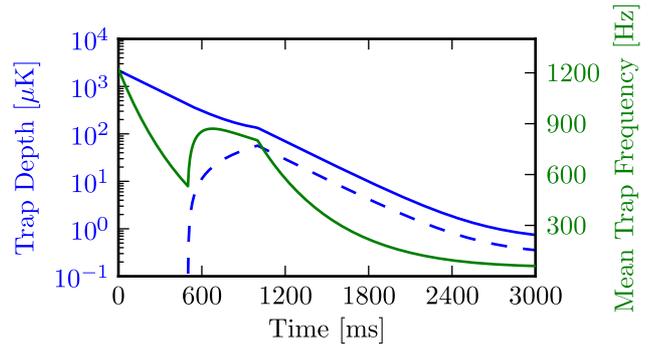}
\caption{Trap depth and mean trapping frequency ($\overline{\omega}/2\pi$) during the evaporation cycle. The solid blue line is the combined trap depth of both beams and the dashed line is the auxiliary beam only. The green line is the mean trapping frequency of the combined beams.}
\label{evap}
\end{figure}

\begin{figure*} [t]
\centering
\includegraphics[width=\textwidth]{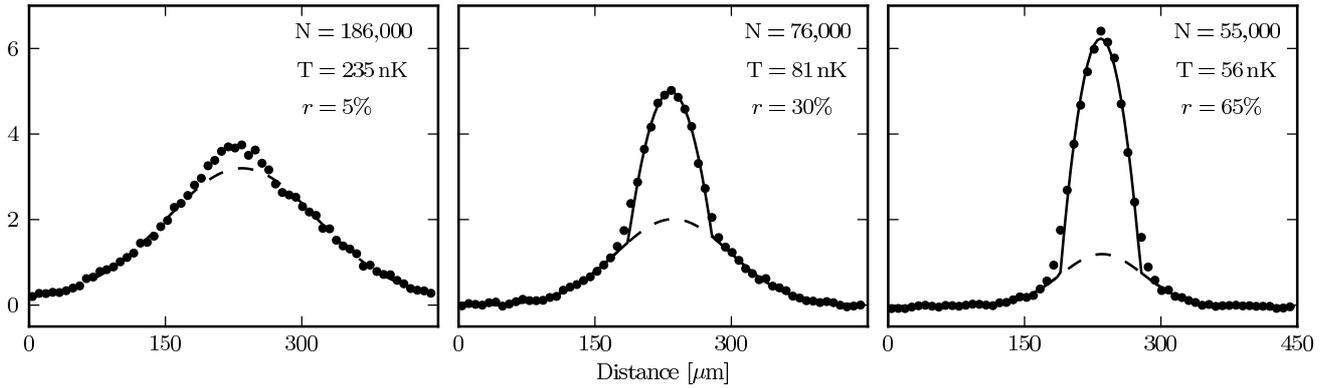}
\caption{Density profiles from thin slices of absorption images after $20 \ms$ drop time. The dashed lines are Gaussian fits to the thermal components. The solid lines are a bimodal fit with the additional Thomas-Fermi profile. Displayed in frame are the total number of atoms (N), the temperature of the thermal components (T), and the condensate fraction (r).}
\label{becprofiles}
\end{figure*}

  The primary beam is on during both the MOT and loading sequence.  For loading the dipole trap, the MOT is compressed to densities of $\approx10^{11} \dens$ by quickly increasing the magnetic field gradient and sweeping the cooling beams to a further detuning \cite{petrich1994}. Specifically, the magnetic field is ramped up to a field gradient of $23\Bz$ in 5~ms and then the detuning of the cooling beams is linearly ramped from $-17\MHz$ to $-120\MHz$ over $40\ms$. The repump intensity is reduced to $\approx 20 \uWcm$ at the start of the detuning sweep.  After the sweep the repump beam is shuttered before the cooling light to ensure the atoms are pumped into the $F=1$ ground state manifold. By standard absorption imaging, we measure $\approx 4$ million atoms in the single beam trap 50 ms after the MOT has been released.

  At 15~W, the single beam has a trap depth of $2\mK$, radial trapping frequencies $\omega_{\mathrm{r}} = \rad 5.6 (0.2) \kHz$ and axial trapping frequency $\omega_{\mathrm{ax}} = \rad 51 (5) \Hz$. The trapping frequencies were measured at several powers via parametric heating and agree well with the expected values for a $25\um$ waist.  If the beam is held at constant power after loading, the temperature decreases due to evaporation until settling at $200\uK$ after 1~s. Here the phase space density peaks at $\approx 2\E{-5}$ with $1.7\E{6}$ atoms remaining. At longer times ($2-10 \s$), the atom number decays with a lifetime of $ 6 \s$. This lifetime is background pressure limited by operation of the rubidium source, as when the source is turned off lifetimes exceeding $10\s$ are observed.

 We typically start reducing the optical power at $50\ms$ after loading, not waiting until initial distribution is fully settled.  If we perform a forced evaporation in the single beam only, the phase space density peaks at $\approx 2\E{-2}$ due to decreasing elastic collision rate as the trap relaxes. The auxiliary beam is thus necessary to compress the atoms, bringing the collision rate back up so evaporation can continue.  Figure~\ref{evap} shows the combined trap depth and mean trapping frequency of both beams during the evaporation cycle.  We reach the critical temperature at $T_{\mathrm{c}} \approx 250 \nK$ with $2.0\E{5}$ atoms remaining. Figure~\ref{becprofiles} shows density profiles of the atom cloud at three points past T$_{\mathrm{c}}$.   The final image has $3.5\E4$ atoms in the condensed phase. The entire evaporation time is $3\s$ and the experiment cycle time is $12\s$ to allow for sufficient MOT loading.

Note that the auxiliary beam is deliberately made to be highly elliptical in order to make the trap anisotropic. In a previous iteration of the experiment, all trap frequencies were nearly identical and with our imaging resolution of $18\um$ we could not see a clearly bimodal distribution. By making the trap anisotropic, the bimodal distribution becomes much more pronounced. Elliptical beam geometries also allow for tailoring favorable crossed beam trapping frequencies without compromising trapping volume and depth.

In summary, we have presented a simple method of reaching condensation by loading one tight confining beam directly from the MOT with the help of one additional crossed beam during evaporation. We have also highlighted difficulties implementing trapping geometries directly with a $1.06\um$ fiber laser which have proven successful using a $10.6\um$ CO$_2$ laser. Our method overcomes the obstacles discussed without adding further complexity to the experimental setup. 

\section*{Acknowledgements}
We acknowledge the support of this work by the National Research Foundation \& Ministry of Education in Singapore. This work was supported by AStar under project No. SERC 052 123 0088.

\bibliographystyle{model1-num-names}

\bibliography{mybib}{}

\begin{thebibliography}{16}
\expandafter\ifx\csname natexlab\endcsname\relax\def\natexlab#1{#1}\fi
\providecommand{\bibinfo}[2]{#2}
\ifx\xfnm\relax \def\xfnm[#1]{\unskip,\space#1}\fi
%Type = Article
\bibitem[{Anderson et~al.(1995)Anderson, Ensher, Matthews, Wieman, and
  Cornell}]{anderson1995}
\bibinfo{author}{M.~Anderson}, \bibinfo{author}{J.~Ensher},
  \bibinfo{author}{M.~Matthews}, \bibinfo{author}{C.~Wieman},
  \bibinfo{author}{E.~Cornell},
\newblock \bibinfo{journal}{Science} \bibinfo{volume}{269}
  (\bibinfo{year}{1995}) \bibinfo{pages}{198--198}.
%Type = Article
\bibitem[{Davis et~al.(1995)Davis, Mewes, Andrews, Van~Druten, Durfee, Kurn,
  and Ketterle}]{davis1995}
\bibinfo{author}{K.~B. Davis}, \bibinfo{author}{M.~O. Mewes},
  \bibinfo{author}{M.~R. Andrews}, \bibinfo{author}{N.~J. Van~Druten},
  \bibinfo{author}{D.~S. Durfee}, \bibinfo{author}{D.~M. Kurn},
  \bibinfo{author}{W.~Ketterle},
\newblock \bibinfo{journal}{Phys. Rev. Lett.} \bibinfo{volume}{75}
  (\bibinfo{year}{1995}) \bibinfo{pages}{3969--3973}.
%Type = Article
\bibitem[{Barrett et~al.(2001)Barrett, Sauer, and Chapman}]{barrett2001}
\bibinfo{author}{M.~Barrett}, \bibinfo{author}{J.~Sauer},
  \bibinfo{author}{M.~Chapman},
\newblock \bibinfo{journal}{Phys. Rev. Lett.} \bibinfo{volume}{87}
  (\bibinfo{year}{2001}) \bibinfo{pages}{10404}.
%Type = Article
\bibitem[{Cennini et~al.(2003)Cennini, Ritt, Geckeler, and Weitz}]{cennini2003}
\bibinfo{author}{G.~Cennini}, \bibinfo{author}{G.~Ritt},
  \bibinfo{author}{C.~Geckeler}, \bibinfo{author}{M.~Weitz},
\newblock \bibinfo{journal}{App. Phys. B} \bibinfo{volume}{77}
  (\bibinfo{year}{2003}) \bibinfo{pages}{773--779}.
%Type = Article
\bibitem[{Weber et~al.(2003)Weber, Herbig, Mark, N\"{a}gerl, and
  Grimm}]{weber2003}
\bibinfo{author}{T.~Weber}, \bibinfo{author}{J.~Herbig},
  \bibinfo{author}{M.~Mark}, \bibinfo{author}{H.~C. N\"{a}gerl},
  \bibinfo{author}{R.~Grimm},
\newblock \bibinfo{journal}{Science} \bibinfo{volume}{299}
  (\bibinfo{year}{2003}) \bibinfo{pages}{232--235}.
%Type = Article
\bibitem[{Dumke et~al.(2006)Dumke, Johanning, Gomez, Weinstein, and
  Jones}]{dumke2006}
\bibinfo{author}{R.~Dumke}, \bibinfo{author}{M.~Johanning},
  \bibinfo{author}{E.~Gomez}, \bibinfo{author}{J.~D. Weinstein},
  \bibinfo{author}{K.~M. Jones},
\newblock \bibinfo{journal}{New J. of Phys.} \bibinfo{volume}{8}
  (\bibinfo{year}{2006}) \bibinfo{pages}{64}.
%Type = Article
\bibitem[{Kinoshita et~al.(2005)Kinoshita, Wenger, and Weiss}]{kinoshita2005}
\bibinfo{author}{T.~Kinoshita}, \bibinfo{author}{T.~Wenger},
  \bibinfo{author}{D.~S. Weiss},
\newblock \bibinfo{journal}{Phys. Rev. A} \bibinfo{volume}{71}
  (\bibinfo{year}{2005}) \bibinfo{pages}{11602}.
%Type = Article
\bibitem[{Couvert et~al.(2008)Couvert, Jeppesen, Kawalec, Reinaudi, Mathevet,
  and Guery-Odelin}]{couvert2008}
\bibinfo{author}{A.~Couvert}, \bibinfo{author}{M.~Jeppesen},
  \bibinfo{author}{T.~Kawalec}, \bibinfo{author}{G.~Reinaudi},
  \bibinfo{author}{R.~Mathevet}, \bibinfo{author}{D.~Guery-Odelin},
\newblock \bibinfo{journal}{Europhysics Lett.)} \bibinfo{volume}{83}
  (\bibinfo{year}{2008}) \bibinfo{pages}{50001}.
%Type = Article
\bibitem[{Cl{\'e}ment et~al.(2009)Cl{\'e}ment, Brantut, Robert-de
  Saint-Vincent, Nyman, Aspect, Bourdel, and Bouyer}]{clement2009}
\bibinfo{author}{J.~F. Cl{\'e}ment}, \bibinfo{author}{J.~P. Brantut},
  \bibinfo{author}{M.~Robert-de Saint-Vincent}, \bibinfo{author}{R.~A. Nyman},
  \bibinfo{author}{A.~Aspect}, \bibinfo{author}{T.~Bourdel},
  \bibinfo{author}{P.~Bouyer},
\newblock \bibinfo{journal}{Phys. Rev. A} \bibinfo{volume}{79}
  (\bibinfo{year}{2009}) \bibinfo{pages}{61406}.
%Type = Article
\bibitem[{Hung et~al.(2008)Hung, Zhang, Gemelke, and Chin}]{hung2008}
\bibinfo{author}{C.~Hung}, \bibinfo{author}{X.~Zhang},
  \bibinfo{author}{N.~Gemelke}, \bibinfo{author}{C.~Chin},
\newblock \bibinfo{journal}{Phys. Rev. A} \bibinfo{volume}{78}
  (\bibinfo{year}{2008}) \bibinfo{pages}{11604}.
%Type = Article
\bibitem[{Beaufils et~al.(2008)Beaufils, Chicireanu, Zanon, Laburthe-Tolra,
  Mar{\'e}chal, Vernac, Keller, and Gorceix}]{beaufils2008}
\bibinfo{author}{Q.~Beaufils}, \bibinfo{author}{R.~Chicireanu},
  \bibinfo{author}{T.~Zanon}, \bibinfo{author}{B.~Laburthe-Tolra},
  \bibinfo{author}{E.~Mar{\'e}chal}, \bibinfo{author}{L.~Vernac},
  \bibinfo{author}{J.~Keller}, \bibinfo{author}{O.~Gorceix},
\newblock \bibinfo{journal}{Phys. Rev. A} \bibinfo{volume}{77}
  (\bibinfo{year}{2008}) \bibinfo{pages}{61601}.
%Type = Article
\bibitem[{Onofrio and Presilla(2002)}]{onofrio2002}
\bibinfo{author}{R.~Onofrio}, \bibinfo{author}{C.~Presilla},
\newblock \bibinfo{journal}{Phys. Rev. Lett.} \bibinfo{volume}{89}
  (\bibinfo{year}{2002}) \bibinfo{pages}{100401}.
%Type = Article
\bibitem[{Adams et~al.(1995)Adams, Lee, Davidson, Kasevich, and
  Chu}]{adams1995}
\bibinfo{author}{C.~S. Adams}, \bibinfo{author}{H.~J. Lee},
  \bibinfo{author}{N.~Davidson}, \bibinfo{author}{M.~Kasevich},
  \bibinfo{author}{S.~Chu},
\newblock \bibinfo{journal}{Phys. Rev. Lett} \bibinfo{volume}{74}
  (\bibinfo{year}{1995}) \bibinfo{pages}{3577--3580}.
%Type = Phdthesis
\bibitem[{Barrett(2002)}]{mythesis}
\bibinfo{author}{M.~D. Barrett}, Ph.D. thesis, Georgia Institute of Techology,
  \bibinfo{year}{2002}.
%Type = Article
\bibitem[{Burt et~al.(1997)Burt, Ghrist, Myatt, Holland, Cornell, and
  Wieman}]{burt1997}
\bibinfo{author}{E.~A. Burt}, \bibinfo{author}{R.~W. Ghrist},
  \bibinfo{author}{C.~J. Myatt}, \bibinfo{author}{M.~J. Holland},
  \bibinfo{author}{E.~A. Cornell}, \bibinfo{author}{C.~E. Wieman},
\newblock \bibinfo{journal}{Phys. Rev. Lett.} \bibinfo{volume}{79}
  (\bibinfo{year}{1997}) \bibinfo{pages}{337--340}.
%Type = Article
\bibitem[{Petrich et~al.(1994)Petrich, Anderson, Ensher, and
  Cornell}]{petrich1994}
\bibinfo{author}{W.~Petrich}, \bibinfo{author}{M.~H. Anderson},
  \bibinfo{author}{J.~E. Ensher}, \bibinfo{author}{E.~A. Cornell},
\newblock \bibinfo{journal}{J. of the Opt. Soc. of America B}
  \bibinfo{volume}{11} (\bibinfo{year}{1994}) \bibinfo{pages}{1332--1335}.

\end{thebibliography}

%\begin{thebibliography}{16}

%\end{thebibliography}

\end{document}